\documentstyle[12pt,epsfig,rotating]{article}
\begin{document}
\baselineskip=16pt

\begin{flushright}
LA-UR-2070
\end{flushright}

\centerline{\bf Oscillations in the GSI electron capture experiment}
\vspace*{0.1in}
\centerline{H. Burkhardt, J. Lowe and G.J. Stephenson Jr. }
\centerline{{\it Physics and Astronomy Department, University of
New Mexico, Albuquerque, \\NM 87131, USA} }
\vspace*{0.05in}
\centerline{T. Goldman}
\centerline{{\it Theoretical Division, Los Alamos National Laboratory,
Los Alamos, \\NM 87545, USA}}
\vspace*{0.05in}
\centerline{Bruce H.J. McKellar}
\centerline{{\it School of Physics, University of Melbourne, Victoria 3010, 
Australia}}
\vspace*{0.05in}
\centerline{31 March 2008}

\vspace{0.1in}

\noindent {\bf Abstract:} In a recent paper, oscillations observed in the
electron capture probability were attributed to the mixing of neutrino
mass eigenstates. This paper is shown to be in error in two respects.

\vspace{0.2in}

In a recent measurement\cite{ref:gsi} at GSI of the electron capture rate 
for hydrogen-like $^{140}Pr$ ions,

\begin{eqnarray}
^{140}Pr^{58+}\rightarrow ^{140}Ce^{58+}+\nu
\label{eq.1}
\end{eqnarray}

\noindent the decaying ions were found to have a non-exponential
time dependence. The approximately sinusoidal modulation superimposed
on the exponential decay was tentatively interpreted as resulting from
the oscillations of the recoil neutrino due to the presence of  at
least two mass eigenstates in the electron neutrino state.

Two papers\cite{ref:ivanov},\cite{ref:faber} have claimed to provide a
theoretical explanation of this effect. We examine here the first of
these and find that it has two errors which, when corrected, remove the
predicted oscillations.

The two errors in the paper by Ivanov {\it et al}.\cite{ref:ivanov} are
the following.

\noindent (1) Ivanov {\it et al}.\cite{ref:ivanov} write the amplitude
for the electron capture as

\begin{eqnarray}
A(^{140}Pr^{58+}\rightarrow ^{140}Ce^{58+}+\nu)(t)=
-i\Sigma_{j=1}^{3}\int_0^t\langle ^{140}Ce^{58+}\nu\mid H_W(\tau)\mid
^{140}Pr^{58+}\rangle d\tau
\label{eq.2}
\end{eqnarray}

Thus the final neutrino state in the amplitude is a sum over neutrino mass
eigenstates.  However, when the neutrino is not observed, one should sum
the probabilities, not the amplitudes, over a complete set of neutrino states. 
This can be done over mass eigenstates, flavour eigenstates, or even the 
states $|\nu_1 + \nu_2 + \nu_3 \rangle/\sqrt{3}$,  $|\nu_1 - \nu_2 \rangle/\sqrt{2}$, 
and  $|\nu_1 + \nu_2 -2 \nu_3 \rangle/\sqrt{6}$, the first of which is the correctly 
normalised version of the state used by Ivanov {\it et al}. The result is the same 
whichever complete set of states is used --- there are no oscillations.

This is just another way of saying that the electron capture process always 
produces only the combination of mass eigenstates corresponding to an 
electron neutrino.

This point is essentially the same as that made by Giunti\cite{ref:giunti}
in a recent comment, who shows that when the correct final state is used,
there are no  oscillations.

\vspace{0.1in}

\noindent (2) A further error in Ivanov {\it et al}.\cite{ref:ivanov}
arises in the treatment of wave packeting. It is essential to include
wave packet structure in the treatment, partly to provide a realistic
description of the experimental set-up but also because oscillations
cannot be observed without wave packeting. The latter point requires
that the spacial width of a packet should be less than the wavelength
of the oscillations.

Ivanov {\it et al}. therefore define the initial state as a wave packet,
so that there is a spread of energies in the inital state. This approach
appears to give difficulties in keeping track of 4-momentum conservation
in the decay. Indeed, Ivanov {\it et al}. appear to conclude that 
4-momentum is not conserved as a result of the spread required 
to produce a wave packet.

An alternative approach is to calculate for a plane-wave initial state,  with
a sharply defined 4-momentum. Then the wave-packeting is introduced by
summing a set of such solutions to produce the required packet size. We 
used this procedure in a paper treating neutral kaon oscillations\cite{ref:usplb1} 
and also in two more general papers that additionally include neutrino 
oscillations from pion decay\cite{ref:usprd},\cite{ref:usplb2}. In this way, 
it is straightforward to include 4-momentum conservation exactly at all stages 
of the calculation, which is desirable because the decay interaction conserves 
4-momentum.  When this is done, the usual result follows simply, namely that 
the particle with mixed mass eigenstates shows oscillations when it is observed, 
but the particle recoiling against it does not.

It is important to treat the kinematics exactly, and not to ignore
energy terms that are comparable with the mass difference between
mass eigenstates or the width of the wave packet. In the course of
our work, we showed that no approximation is necessary in the
evaluation; in Ref.\cite{ref:usprd} we derived the oscillation
expression exactly, to all orders and including the correct kinematics.

The situation proposed by Ivanov {\it et al}.\cite{ref:ivanov} is
strikingly similar to that described in a series of papers by
Widom and collaborators\cite{ref:widom} on the process
$\pi {\rm}p\rightarrow \Lambda K^0$, in which it was predicted
that oscillations should be observed in {\it both} the $K^0$ {\it and}
the $\Lambda$ distributions (which is certainly not the case, 
experimentally) and that the usual expression relating the oscillation 
frequency to $m_L-m_S$ is in error. On examining the algebra of 
Ref. \cite{ref:widom}, we found\cite{ref:usplb1},\cite{ref:usprd},\cite{ref:usplb2} 
a kinematic approximation in the evaluation which gave rise to these 
unexpected effects, which are quite inconsistent with experiment. With 
the exact treatment, these unexpected effects disappeared and the standard
result was recovered. (For yet another exact momentum conservation
treatment using plane waves, see Goldman\cite{ref:goldman}.)

Finally, we recall that Aharonov and Moinester\cite{ref:Aharanov} examined
this question long ago in connection with a pion decay experiment at the 
Los Alamos Meson Physics Facility in the early 1980's. Initially, there 
appeared to be oscillations about exponential decay registered in the 
appearance of muons in the final state. This was eventually traced to an 
experimental problem involving cross-talk between timing cables\cite{ref:Moinester}. 
Aharonov and Moinester also concluded that these 
oscillations could not have had a basis in neutrino physics, due to the 
orthogonality of the neutrino mass eigenstates\cite{ref:Aharanov}.

We conclude that the papers of Refs.\cite{ref:ivanov} and \cite{ref:faber}
do not provide an explanation of the oscillations observed in the GSI
experiment\cite{ref:gsi} in terms of neutrino mixing and that the 
experimental results are not a consequence of neutrino oscillations.


\vspace{0.5cm}

\begin{thebibliography}{99}


\bibitem{ref:gsi}
Yu.A. Litvinov {\it et al}., arxiv/0711.3709; 
{\it Phys.\ Rev.\ Lett.\ }{\bf 99}, 262501 (2007).

\bibitem{ref:ivanov}
A.N. Ivanov {\it et al}., arxiv/0801.2121.

\bibitem{ref:faber}
M. Faber, arxiv/0711.3262.

\bibitem{ref:giunti}
Carlo Giunti, arxiv/0801.4639. For a reply, see A.N. Ivanov {\it et al}., 
arxiv/0803.1289.

\bibitem{ref:usplb1}
J. Lowe, B. Bassalleck, H. Burkhardt, A. Rusek, G.J. Stephenson Jr.
and T. Goldman,
hep-ph/9605234; {\it Phys.\ Lett.\ }{\bf B384}, 288 (1996).

\bibitem{ref:usprd}
H. Burkhardt, T. Goldman, G.J. Stephenson Jr. and J. Lowe,
{\it Phys.\ Rev.\ D}{\bf 59}, 054018 (1999).

\bibitem{ref:usplb2}
H. Burkhardt, T. Goldman, G.J. Stephenson Jr. and J. Lowe,
hep-ph/0302084; {\it Phys.\ Lett.\ }{\bf B566}, 137 (2003).

\bibitem{ref:widom}
Y.N. Srivastava, A. Widom and E. Sassaroli,
{\it Phys.\ Lett.\ }{\bf B344}, 436 (1995);
Y.N. Srivastava, A. Widom and E. Sassaroli,
{\it Zeit.\ Phys.\ }{\bf C66}, 601 (1995).

\bibitem{ref:goldman} T. Goldman, unpublished, hep-ph/9604357.

\bibitem{ref:Aharanov} Y. Aharanov and M.A. Moinester, unpublished.

\bibitem{ref:Moinester} M.A. Moinester, private communication.



\end{thebibliography}
\end{document}